\RequirePackage{fix-cm}
\documentclass[twocolumn]{svjour3}          
\smartqed  
\usepackage{graphicx}
\usepackage{mathptmx}      
\usepackage{braket}
\usepackage{cite}
\usepackage[utf8]{inputenc}
\usepackage{amssymb}
\usepackage{amsmath}
\usepackage{hyperref}
\usepackage{mathdots}
\usepackage{placeins}
\usepackage{bm}
\usepackage{subfigure}
\journalname{myjournal}

\begin{document}

\title{Analysis and synthesis of feature map for kernel-based quantum classifier
	\thanks{Y. Suzuki, H. Yano: Equally contributing authors.}
}


\author{Yudai Suzuki \and
	Hiroshi Yano \and
	Qi Gao \and
	Shumpei Uno \and
	Tomoki Tanaka \and
	Manato Akiyama \and
	Naoki Yamamoto
}


\institute{
	Yudai Suzuki
	\at Department of Mechanical Engineering, Keio University, Hiyoshi 3-14-1, Kohoku,
	Yokohama 223-8522, Japan
	\and
	Hiroshi Yano
	\at Department of Information and Computer Science, Keio University, Hiyoshi 3-14-1,
	Kohoku, Yokohama 223-8522, Japan
	\and
	Qi Gao \and Shumpei Uno \and Tomoki Tanaka \and Naoki Yamamoto
	\at Quantum Computing Center, Keio University, Hiyoshi 3-14-1, Kohoku,
	Yokohama 223-8522, Japan
	\and
	Manato Akiyama
	\at Department of Biosciences and Informatics, Keio University, Hiyoshi 3-14-1,
	Kohoku, Yokohama 223-8522, Japan
	\and
	Qi Gao
	\at Mitsubishi Chemical Corporation Science \& Innovation Center, 1000,
	Kamoshida-cho, Aoba-ku, Yokohama 227-8502, Japan
	\and
	Shumpei Uno
	\at Mizuho Information \& Research Institute, Inc., 2-3 Kanda-Nishikicho, Chiyoda-ku,
	Tokyo 101-8443, Japan
	\and
	Tomoki Tanaka
	\at Mitsubishi UFJ Financial Group, Inc. and MUFG Bank, Ltd., 2-7-1 Marunouchi,
	Chiyoda-ku, Tokyo 100-8388, Japan
	\and
	Naoki Yamamoto
       \at
       Department of Applied Physics and Physico-Informatics, Keio University, Hiyoshi 3-14-1,
       Kohoku, Yokohama 223- 8522, Japan,
       \email{yamamoto@appi.keio.ac.jp}
}

\date{Received: date / Accepted: date}

\maketitle


\begin{abstract}
A method for analyzing the feature map for the kernel-based quantum classifier is
developed;
that is, we give a general formula for computing a lower bound of the exact training
accuracy, which helps us to see whether the selected feature map is suitable for linearly
separating the dataset.
We show a proof of concept demonstration of this method for a class of 2-qubit
classifier, with several 2-dimensional dataset.
Also, a synthesis method, that combines different kernels to construct a better-performing
feature map in a lager feature space, is presented.
\end{abstract}



\section{Introduction\label{intro}}

Over the last 20 years, the unprecedented improvements in the cost-effectiveness ration
of computer, together with improved computational techniques, make machine learning
widely applicable in every aspect of our lives such as education, healthcare, games, finance,
transportation, energy, business, science and engineering~\cite{HastieTF09,Alpaydin16}.
Among numerous developed machine learning methods, Support Vector Machine (SVM)
is a very established one which has become an overwhelmingly popular choice for data
analysis~\cite{Boser1992}.
In SVM method, a nonlinear dataset is transformed via a feature map to another dataset
and is separated by a hyperplane in the feature space, which can be effectively performed
using the kernel trick.
In particular, the Gaussian kernel is often used.

Quantum computing is expected to speed-up the performance of machine learning
through exploiting quantum mechanical properties including superposition, interference,
and entanglement.
As for the quantum classifier, a renaissance began in the past few years, with the
quantum SVM method \cite{Rebentrost2014}, the quantum circuit learning method
\cite{Mitarai2018}, and some parallel development by other research groups
\cite{Neven 2018,Zhuang Zhang 2019,Wilson2018}.
In particular, the kernel method can be exploited as a powerful mean that can be
introduced in the quantum SVM as in the classical case \cite{Rebentrost2014,Chatterjee 2017,Bishwas2018,LiTongyang2019,havlivcek2019,Schuld2019,Nori 2019,Park Rhee 2019,Negoro 2019,Lloyd2020,LaRose2020};
importantly, the concept of kernel-based quantum classifier has been experimentally
demonstrated on a real device~\cite{havlivcek2019,Park Rhee 2019,Negoro 2019}.
These studies show the possibility that machine learning will get a further boost by using
quantum computers in the near future.

For all the developed quantum classification methods, the feature map plays a role to
encode the dataset taken from its original low dimensional real space onto a high
dimensional quantum state space (i.e., the Hilbert space).
A possible advantage of quantum classifier lies in the fact that this high-dimensional
feature space is realized on a physical quantum computer even with a medium number
of qubits, as well as the fact that the kernel could be computed faster than the classical
case.
However, in the framework of using gate-based quantum computers, a suitable feature
map has to be explicitly specified, which is of course less trivial than specifying a suitable
kernel.
(Note that a kernel induces a feature map through the reproducing kernel Hilbert space.)
Actually, for choosing a suitable feature map, one could prepare many map candidates and
try to find a best one by comparing the results of training accuracy attained with all those
maps, but this clearly needs numerous times of classification or regression analysis.
Hence it is desirable if we have a method for easily having a rough estimate of the training
accuracy of every feature map candidate.

This paper gives one such method, based on the {\it minimum accuracy}, a lower bound of
the exact training accuracy attained by any optimized classifier.
The minimum accuracy is determined only from a chosen feature map and the input
classical dataset, hence it can be used to screen a library of suitable feature maps.
A critical drawback of this method is that it needs calculation of the order of the dimension
of the feature space, which is exponential to the number of qubits.
Hence, to show the proof of concept, in this paper we study the case where the quantum
classifier is composed of only two qubits (see Section \ref{sec:conclusion} for a possible
extension of the method).
This simple setup gives us an explicit form of the feature map and further a visualization
of the encoded input data distribution in the feature space;
this eventually enables us to easily calculate the minimum accuracy.
Moreover, the visualized feature map candidates might be exploited for combining them to
construct a better-performing feature map and accordingly a better kernel.
Although the concept of these synthesizing tools is device-independent, in this paper we
demonstrate the idea in the framework of \cite{havlivcek2019}, with a special type of five
encoding functions and four 2-dimensional nonlinear datasets.


\section{Methods}

\subsection{Real-vector representation of the feature map via Pauli decomposition}

In the SVM method with quantum kernel estimator, the feature map transforms an input
dataset to a set of multi-qubit states, which forms the feature space (i.e., Hilbert space);
then the kernel matrix is constructed by calculating all the inner products of quantum
states, and it is finally used in the (standard) SVM for classifying the dataset.
Surely different feature maps lead to different kernels and accordingly influence on the
classification accuracy, meaning that a careful analysis of the feature space is necessary.
However, due to the complicated structure of the feature space, such analysis is in general
not straightforward.
Here, we propose the Pauli-decomposition method for visualizing the feature space, which
might be used as a guide to select a suitable feature map.

We begin with a brief summary of the kernel-based quantum SVM method proposed in
\cite{havlivcek2019}.
First, a $\tilde{n}$-dimensional classical data $\bm{x}\in\mathbb{R}^{\tilde{n}}$ is encoded
into the unitary operator $\mathcal{U}_{\Phi(\bm{x})}$ through an {\it encoding function}
${\Phi(\bm{x})}$, and it is applied to the initial state $\ket{0}^{\otimes n}$ with $\ket{0}$ the
qubit ground state.
Thus, the feature map is a transformation from the classical data $\bm{x}$ to the
quantum state $\ket{\Phi(\bm{x})}=\mathcal{U}_{\Phi(\bm{x})}\ket{0}^{\otimes n}$, and the
feature space is $(\mathbb{C}^2)^{\otimes n}=\mathbb{C}^{2^n}$.
The kernel is then naturally defined as
$K(\bm{x},\bm{z})=|\braket{\Phi(\bm{x})|\Phi(\bm{z})}|^2$;
this quantity can be practically calculated as the ratio of zero strings $0^n$ in the $Z$-basis
measurement result, for the state
$\mathcal{U}^\dagger_{\Phi(\bm{x})} \mathcal{U}_{\Phi(\bm{z})} \ket{0}^{\otimes n}$.
Finally, the constructed kernel is used in the standard manner in SVM;
that is, a test data $\bm{x}$ is classified into two categories depending on the sign of
\begin{equation}
         \sum_{i=1}^N \alpha_i y_i K(\bm{x}_i,\bm{x}) + b,
\end{equation}
where $(\bm{x}_i, y_i)~(i=1,\ldots, N)$ are the pairs of training data, and $(\alpha_i, b)$ are
the optimized parameters.

Now we introduce the real-vector representation of the feature map.
The key idea is simply to use the fact that the kernel can be expressed in terms of the
density operator $\rho(x) = \ket{\Phi(\bm{x})}\bra{\Phi(\bm{x})}$ as
\begin{equation}
         K(\bm{x},\bm{z})
           =|\braket{\Phi(\bm{x})|\Phi(\bm{z})}|^2
           = \mathrm{tr}\left[ \rho(\bm{x}) \rho(\bm{z}) \right],
\label{eq:trace_rho}
\end{equation}
and the density operator can be always expanded by the set of Pauli operators as
\begin{equation}
      \rho(\bm{x}) = \sum_{i=1}^{4^n} a_i(\bm{x}) \sigma_i
\label{eq:paulidec}
\end{equation}
with $a_i(\bm{x})\in \mathbb{R}$ and $\sigma_i \in P_n=\{I, X, Y, Z\}^{\otimes n}$
the multi-qubit Pauli operators.
The followings are examples of the elements of $P_2$:
\begin{equation*}
  \begin{split}
    XI &= \left[\begin{array}{rr}
                0 & 1  \\
                1 & 0  \\
             \end{array}\right]
             \otimes
             \left[\begin{array}{rr}
                1 & 0  \\
                0 & 1  \\
             \end{array}\right], ~~
      ZY = \left[\begin{array}{rr}
                1 & 0  \\
                0 & -1  \\
             \end{array}\right]
             \otimes
             \left[\begin{array}{rr}
                0 & -i  \\
                i & 0  \\
             \end{array}\right], ~~ \\
             YZ &= \left[\begin{array}{rr}
                0 & -i  \\
                i & 0  \\
             \end{array}\right]
             \otimes
             \left[\begin{array}{rr}
                1 & 0  \\
                0 & -1 \\
             \end{array}\right].
  \end{split}
\end{equation*}
Then, by substituting Eq.~\eqref{eq:paulidec} into Eq.~\eqref{eq:trace_rho} and using
the trace relation $\mathrm{tr}\left( \sigma_i \sigma_j \right)= 2^n \delta_{i,j}$, the kernel
can be written as
\begin{equation}
     K(\bm{x}, \bm{z}) = 2^n \sum_{i = 1}^{4^n} a_{i}(\bm{x})a_{i}(\bm{z}),
\label{eq:p_coef_prod}
\end{equation}
meaning that the vector $\bm{a}(\bm{x})=[a_1(\bm{x}), \ldots, a_{4^n}(\bm{x})]^\top$
serves as the feature map corresponding to the kernel $K(\bm{x},\bm{z})$.
That is, the input dataset $\{\bm{x}_i\}$ are encoded into the set of vectors
$\{ \bm{a}(\bm{x}_i)\}$ in a (bigger) real feature space $\mathbb{R}^{4^n}$ and will be
classified by the SVM with the kernel \eqref{eq:p_coef_prod}.
Note that $\bm{a}(\bm{x})$ is a generalization of the Bloch vector, and thus the
corresponding feature space is interpreted as the generalized Bloch sphere.


\subsection{Feature map for the 2-qubit classifier}
\label{sec:formulation}

\begin{figure}[tb]
\begin{center}
\includegraphics[width=1.0\linewidth]{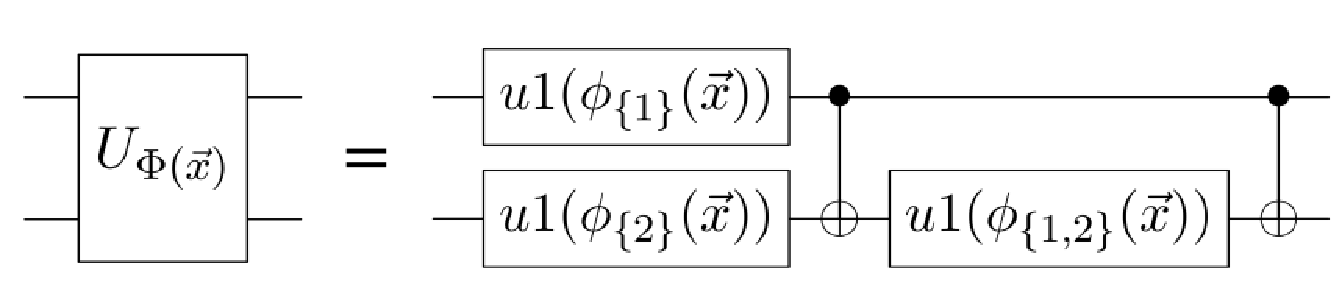}
\caption{Quantum circuit of $U_{\Phi(\bm{x})}$ with the encoding function
$\Phi(\bm{x})=\{ \phi_1(\bm{x}), \phi_2(\bm{x}), \phi_{1,2}(\bm{x})\}$, composed of the
three phase-shift gates of the form $u1(\phi)={\rm diag}\{1, e^{-i\phi} \}$ and
two Controlled-NOT gates.}
\label{fig:circ}
\end{center}
\end{figure}

In this paper, we study the 2-qubit classifier proposed in \cite{havlivcek2019};
an input data $\bm{x}\in\mathbb{R}^{\tilde{n}}$ is mapped to the unitary operator
$\mathcal{U}_{\Phi(\bm{x})}$, which is composed of two layers of Hadamard gate
$H^{\otimes 2}$ and the unitary gate $U_{\Phi(x)}$ as follows:
\begin{equation}
        \mathcal{U}_{\Phi(\bm{x})}=U_{\Phi(\bm{x})}H^{\otimes 2}U_{\Phi(\bm{x})}H^{\otimes 2},
\label{eq:unitary}
\end{equation}
where
\begin{equation}
     U_{\Phi(\bm{x})}
        =\exp\Big(
               i\phi_1(\bm{x})ZI
                + i\phi_2(\bm{x})IZ
                  + i\phi_{1,2}(\bm{x})ZZ
               \Big),
\label{eq:unitarycomponent}
\end{equation}
and $\Phi(\bm{x})=\{ \phi_1(\bm{x}), \phi_2(\bm{x}), \phi_{1,2}(\bm{x})\}$ is the set of
encoding functions.
The quantum circuit representation realizing this unitary gate is shown in Fig.~\ref{fig:circ}.
The three user-defined encoding functions $\phi_1(\bm{x}), \phi_2(\bm{x})$, and
$\phi_{1,2}(\bm{x})$ nonlinearly transform the input data $\bm{x}$ into the qubit
$\ket{\Phi(\bm{x})}=\mathcal{U}_{\Phi(\bm{x})} \ket{0}^{\otimes 2}$.
A lengthy calculation then gives the explicit Pauli decomposed form \eqref{eq:paulidec}
of the density operator $\rho(\bm{x})=\ket{\Phi(\bm{x})}\bra{\Phi(\bm{x})}$;
the coefficients $\{ a_i(\bm{x})\}$ with $i=II, XI, YI, \ldots, ZZ$ are listed in Table~\ref{tb:pdc}.
The coefficients are composed of bunch of trigonometric functions, which make
the kernel complicated enough to transform the input data highly nonlinearly.

\begin{table*}[hbtp]
\begin{center}
\caption{Coefficients of the density operator \eqref{eq:paulidec} in the setup shown in
Section \ref{sec:formulation}; that is, $\{ a_i(\bm{x})\}$ with $i=II, XI, YI, \ldots, ZZ$. }
\label{tb:pdc}
\begin{tabular}{c|c} \hline
Index $i$ & Pauli decomposition coefficients $a_i$ \\ \cline{1-2}
II & $1/4$ \\
XI & $\{\sin{\phi_{1}}(\sin{\phi_{2}}\sin{\phi_{1,2}}^{2}+\sin{\phi_{1}}\cos{\phi_{1,2}}^{2}+\cos{\phi_{2}}\cos{\phi_{1}}\sin{\phi_{1,2}})\}/4$ \\
YI & $\{-\sin{\phi_{2}}\cos{\phi_{1}}\sin{\phi_{1,2}}^{2}-\sin{\phi_{1}}\cos{\phi_{1}}\cos{\phi_{1,2}}^{2}+\cos{\phi_{2}}\sin{\phi_{1}}^{2}\sin{\phi_{1,2}}\}/4$ \\
ZI & $\cos{\phi_{1}}\cos{\phi_{1,2}}/4$ \\
IX & $\{\sin{\phi_{2}}(\sin{\phi_{1}}\sin{\phi_{1,2}}^{2}+\sin{\phi_{2}}\cos{\phi_{1,2}}^{2}+\cos{\phi_{1}}\cos{\phi_{2}}\sin{\phi_{1,2}})\}/4$ \\
XX & $\{\sin{\phi_{1}}^{2}\sin{\phi_{2}}^{2}+\sin{\phi_{1,2}}\cos{\phi_{1}}\cos{\phi_{2}}(\sin{\phi_{1}}+\sin{\phi_{2}})\}/4$ \\
YX & $\{-\sin{\phi_{2}}^{2}\sin{\phi_{1}}\cos{\phi_{1}}+\sin{\phi_{1,2}}\cos{\phi_{2}}(\sin{\phi_{1}}\sin{\phi_{2}}-\cos{\phi_{1}}^{2})\}/4$ \\
ZX & $\{\cos{\phi_{1,2}}(-\sin{\phi_{1}}\cos{\phi_{2}}\sin{\phi_{1,2}}+\cos{\phi_{1}}\sin{\phi_{2}}^{2}+\sin{\phi_{2}}\cos{\phi_{2}}\sin{\phi_{1,2}})\}/4$ \\
IY & $\{-\sin{\phi_{1}}\cos{\phi_{2}}\sin{\phi_{1,2}}^{2}-\sin{\phi_{2}}\cos{\phi_{2}}\cos{\phi_{1,2}}^{2}+\cos{\phi_{1}}\sin{\phi_{2}}^{2}\sin{\phi_{1,2}}\}/4$ \\
XY & $\{-\sin{\phi_{1}}^{2}\sin{\phi_{2}}\cos{\phi_{2}}+\sin{\phi_{1,2}}\cos{\phi_{1}}(\sin{\phi_{1}}\sin{\phi_{2}}-\cos{\phi_{2}}^{2})\}/4$ \\
YY & $\{\sin{\phi_{1}}\cos{\phi_{1}}\sin{\phi_{2}}\cos{\phi_{2}}-\sin{\phi_{1,2}}(\cos{\phi_{2}}^{2}\sin{\phi_{1}}+\sin{\phi_{2}}\cos{\phi_{1}}^{2})\}/4$ \\
ZY & $\{\sin{\phi_{2}}(-\sin{\phi_{1}}\sin{\phi_{1,2}}\cos{\phi_{1,2}}-\cos{\phi_{2}}\cos{\phi_{1}}\cos{\phi_{1,2}}+\sin{\phi_{2}}\cos{\phi_{1,2}}\sin{\phi_{1,2}})\}/4$ \\
IZ & $\cos{\phi_{2}}\cos{\phi_{1,2}}/4$ \\
XZ & $\{\cos{\phi_{1,2}}(-\sin{\phi_{2}}\cos{\phi_{1}}\sin{\phi_{1,2}}+\cos{\phi_{2}}\sin{\phi_{1}}^{2}+\sin{\phi_{1}}\cos{\phi_{1}}\sin{\phi_{1,2}})\}/4$ \\
YZ & $\{\sin{\phi_{1}}(-\sin{\phi_{2}}\sin{\phi_{1,2}}\cos{\phi_{1,2}}-\cos{\phi_{1}}\cos{\phi_{2}}\cos{\phi_{1,2}}+\sin{\phi_{1}}\cos{\phi_{1,2}}\sin{\phi_{1,2}})\}/4$ \\
ZZ & $\cos{\phi_{1}}\cos{\phi_{2}}/4$ \\
\hline
\end{tabular}
\end{center}
\end{table*}


\subsection{Minimum accuracy}
\label{sec:min accuracy}

\begin{figure}[tb]
\begin{center}
\includegraphics[width=1.0\linewidth]{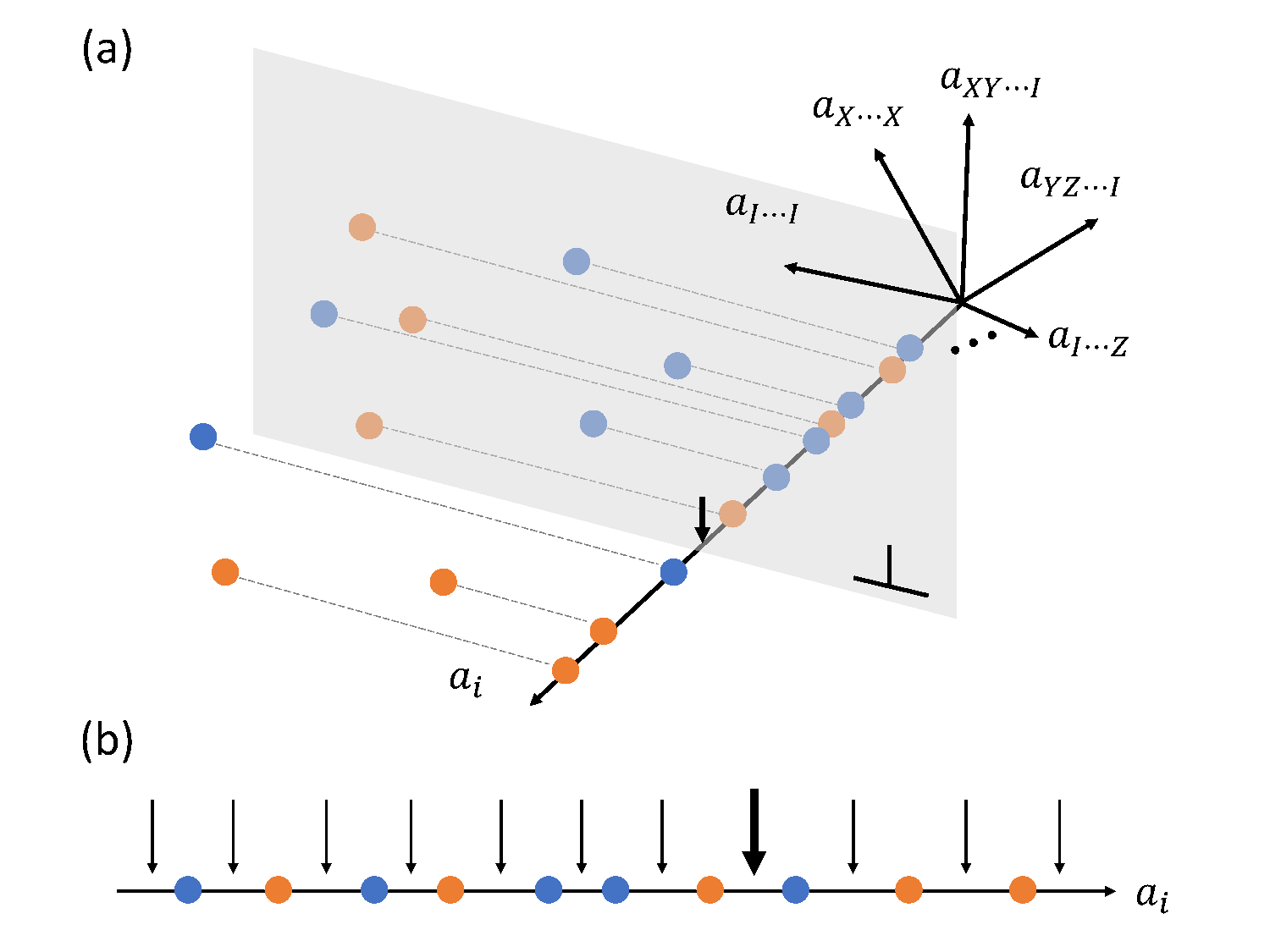}
\caption{
An example for calculating the minimum accuracy in the case of $N=10$ data points
$\{ \bm{a}(\bm{x}_k) \}_{k=1,\ldots,10}$.
The blue and orange points indicate the values of $y_k=-1$ and $y_k=+1$, respectively.
(a) The data points  are projected onto the $a_{i}$-axis in the feature space;
then they are classified at a chosen threshold indicated by the arrow, which is equivalent to
separating the data with the hyperplane orthogonal to the axis.
(b) 1-dimensional view of the data $\{ a_i(\bm{x}_k) \}_{k=1,\ldots, 10}$.
At the same $j=8$th threshold as above, indicated by the solid arrow, we have $N_{-}=4$
and $N_{+}=3$ (the number of blue and orange points in the left side of the arrow, respectively);
then the accuracy \eqref{accuracy at the threshold} is $R_i^8=0.6$.
Changing the place of the threshold (i.e., the index $j$), as indicated by thin arrows, we find
that the maximum of these values is $R_i = \max_j R^{j}_{i}=0.7$.
}
\label{fig:class rate}
\end{center}
\end{figure}

The minimum accuracy is defined as the maximum classification accuracy where the
hyperplane used for classifying the training dataset is restricted to being orthogonal
to any basis axis in the feature space.
The main points of this definition are as follows; due to this restriction, the minimum
accuracy is calculated without respect to the actual classifiers; moreover, it gives a lower
bound of the accuracy for the training dataset achieved by any optimized classifier,
because the optimized hyperplane is not necessarily orthogonal to any basis axis.
This means that the minimum accuracy can be used to evaluate the chosen feature map
and accordingly the kernel, without designing an actual classifier;
in particular, if the minimum accuracy takes a relatively large value, any optimized
classifier is guaranteed to achieve an equal or a higher accuracy in that feature space.
Note that a similar concept is found in Ref.~\cite{aronoff 1985}, which is yet defined
in a different way.

In this study, the feature map is given by the vector
$\bm{a}(\bm{x})=[a_1(\bm{x}), \ldots, a_{4^n}(\bm{x})]^\top$, and the minimum accuracy is
calculated as follows, for the training dataset $\{\bm x_{k}, y_{k}\}_{k=1,\ldots, N}$, for the
case $N$ being an even number (if $N$ is an odd number, generate one more training data).
In particular, we assume that the output $y_k=+1$ has been assigned to $N/2$ data points
and $y_k=-1$ for the rest $N/2$ data points.

\begin{description}

\item[(i)]
For a fixed index $i\in\{1,\cdots,4^n\}$, consider the dataset $\{ a_i(\bm{x}_k) \}_{k=1,\ldots, N}$,
which are the projection of all the transformed data onto the $i$-th axis in the feature space,
as shown in Fig.~\ref{fig:class rate}(a).

\item[(ii)]
Choose a hyperplane orthogonal to this $i$-th axis; they intersect at the threshold between
a pair of neighboring projected data points, as indicated by the thick arrow in
Fig. ~\ref{fig:class rate}(a) and (b).

\item[(iii)]
Calculate the accuracy at the $j$-th threshold as follows.
Let $N_+$ and $N_{-}$ be the number of data points with output $y_k=+1$ and $y_k=-1$
in the left side of the threshold, respectively.
If $N_+ > N_-$, the desirable classification pattern of the dataset is such that the points with
$y_k=+1$ are in the left and the points with $y_k=-1$ are in the right;
now the number of points with $y_k=-1$ is $N/2 - N_-$, meaning that the classification
accuracy is $(N_+ + N/2 - N_-)/N$.
(The perfect case is such that $N_+=N/2$ and $N_-=0$, leading that the accuracy is $1$.)
Combining the case $N_+ < N_-$, hence, the accuracy is defined by
\begin{equation}
\label{accuracy at the threshold}
     R^{j}_{i} = \frac{1}{N} \Big( {\rm max}\{N_+, N_{-} \} + \frac{N}{2} - {\rm min}\{N_+, N_{-}\} \Big).
\end{equation}
Recall that this quantity \eqref{accuracy at the threshold} corresponds to the accuracy
of classifying the dataset by the hyperplane orthogonal to the $i$-th axis at the threshold.

\item[(iv)]
Calculate the accuracy for all the thresholds with indices $j=1,\ldots, N+1$, and then take
the maximum: $R_i = \max_j R^{j}_{i}$.

\item[(v)]
The minimum accuracy is defined as $R = \max_{i} R_i$, where the index runs from $i=1$
to $i=4^n$.

\end{description}
A simple example to demonstrate calculating the minimum accuracy is given in
Fig.~\ref{fig:class rate}.
Note that the above procedure can be readily conducted for the 2-qubit case, using
the explicit form of $\{a_{i}(\bm{x})\}$ listed in Table ~\ref{tb:pdc}.


\section{Results and Discussions}

\subsection{Classification accuracy with different encoding functions}

\begin{figure}[h]
\begin{center}
\includegraphics[width=0.9\linewidth]{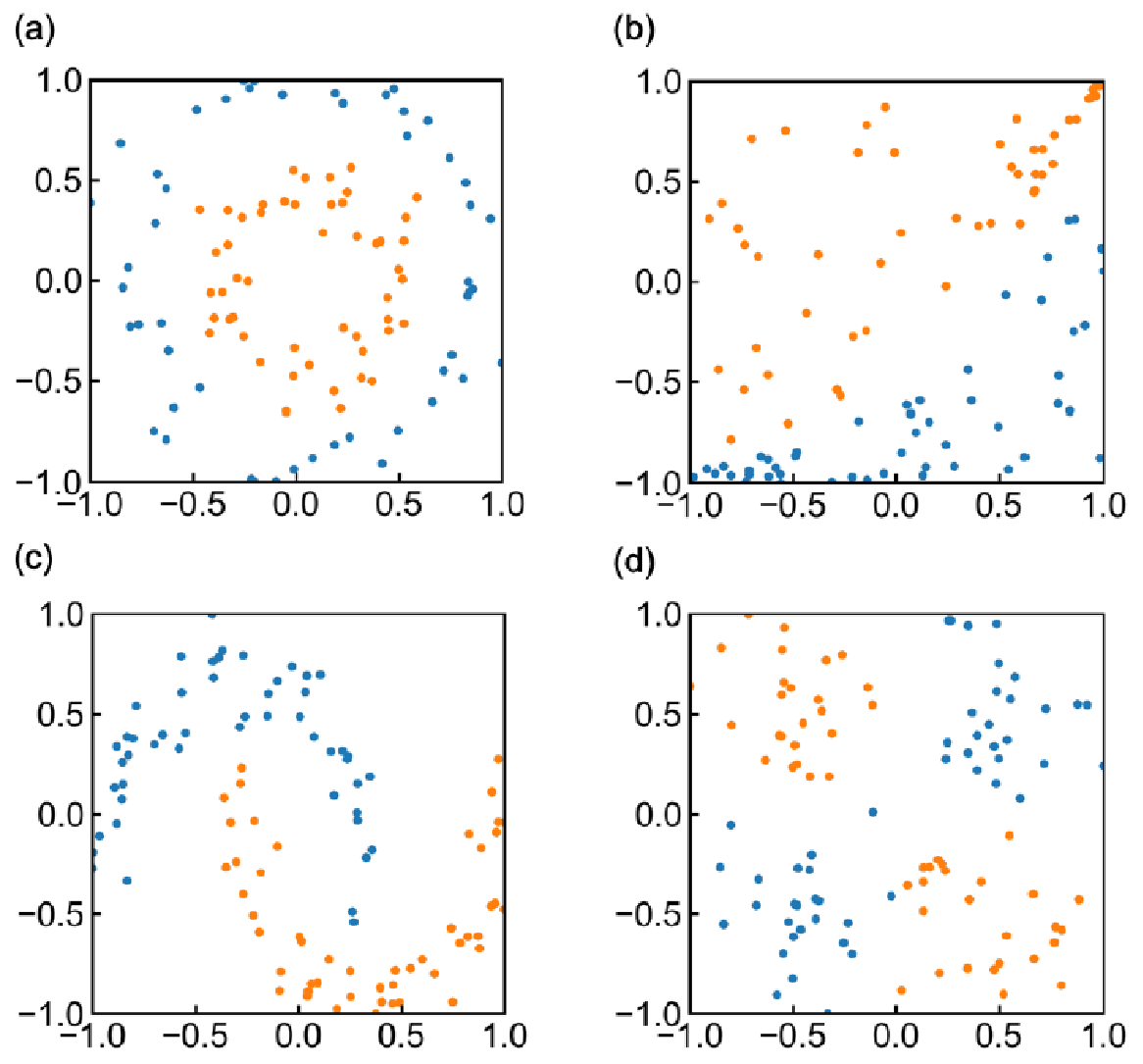}
\caption{Dataset called (a) Circle, (b) Exp, (c) Moon, and (d) Xor.}
\label{fig:dataset}
\end{center}
\end{figure}

Here we apply the quantum SVM method, with several encoding functions, to some
benchmark classification problems.
We consider the nonlinear 2-dimensional datasets named Circle, Exp, Moon, and Xor,
shown in Fig.~\ref{fig:dataset};
in each case, $N=100$ data points $(\bm{x}_k, y_k)~(k=1,\ldots, 100)$ are generated and
categorized to two groups depending on $y_k=+1$ or $y_k=-1$ (orange or blue points
in the figure).
Each dataset is encoded into the 2-qubit quantum state, with the following five
encoding functions:
\begin{align}
       \phi_{\, 1}(\bm{x}) & = x_1, ~ \phi_{\, 2}(\bm{x}) = x_2, ~ \phi_{\, 1,2}(\bm{x}) = \pi x_1 x_2,
       \label{eq:ef1}   \\
       \phi_{\, 1}(\bm{x}) & = x_1, ~ \phi_{\, 2}(\bm{x}) = x_2, ~ \phi_{\, 1,2}(\bm{x})
                  = \frac{\pi}{2}(1 - x_1) (1 - x_2),
       \label{eq:ef2}   \\
       \phi_{\, 1}(\bm{x}) & = x_1, ~ \phi_{\, 2}(\bm{x}) = x_2, ~ \phi_{\, 1,2}(\bm{x})
                  = \exp\left( \frac{|x_1 - x_2|^2}{ 8/\ln(\pi) } \right),
       \label{eq:ef3}    \\
       \phi_{\, 1}(\bm{x}) & = x_1, ~ \phi_{\, 2}(\bm{x}) = x_2, ~ \phi_{\, 1,2}(\bm{x})
                  = \frac{\pi}{3 \cos(x_1) \cos(x_2)},
       \label{eq:ef4} \\
       \phi_{\, 1}(\bm{x}) & = x_1, ~ \phi_{\, 2}(\bm{x}) = x_2, ~ \phi_{\, 1,2}(\bm{x})
                  = \pi  \cos(x_1) \cos(x_2).
       \label{eq:ef5}
\end{align}
The functions $\phi_{\, 1,2}(\bm{x})$ are chosen from a set of various nonlinear functions
in the range of $2\pi$, i.e. ${\rm max}(\phi) - {\rm min}(\phi) \le 2\pi$ for $x_1, x_2 \in [-1,1]$.
In particular, the coefficient of \eqref{eq:ef3} and \eqref{eq:ef4} are determined empirically
so that the resulting classifier achieves a high accuracy on the prepared datasets.
Also the reason of fixing $\phi_{\, 1}(\bm{x})=x_1$ and $\phi_{\, 2}(\bm{x})=x_2$ for all the
encoding functions is that here we aim to investigate the dependence of the classification
accuracy on $\phi_{\, 1,2}(\bm{x})$.
In this work, the classification accuracy are evaluated as the average accuracy of the 5-fold
cross validation, where one dataset is divided into 5 groups with equal number of datasets
(i.e., 20 data-points), for both the training and test dataset.
All the calculations are carried out using QASM simulator included in the Qiskit package
\cite{Qiskit};
to construct each element of the kernel, 10,000 shots (measurements) is performed.
Also to perform the optimization procedure of SVM, scikit-learn, a popular machine learning
library for Python, was employed; in particular the hyperparameter $C$ is set to $10^{10}$
for realizing the hard-margin SVM, which is the scenario where the notion of minimum
accuracy is valid.

The classification accuracy of the four datasets, achieved by the above four encoding
functions, are shown in Table~\ref{tb:train} for the training case and Table~\ref{tb:test} for
the test case.
Overall, the function \eqref{eq:ef4} achieves good accuracy, which is larger than 0.95 for
the training set and 0.88 for the test set.
On the other hand, the function \eqref{eq:ef1} does not always work well for classification;
this function achieves the accuracy 1.00 for the training dataset of Circle and Xor, whereas
the accuracy for the training Moon dataset is decreased to 0.85.
Hence the different encoding functions, which lead to the different feature maps and kernels,
may largely influence the resulting classification accuracy.

\begin{table}
\begin{center}
\caption{Classification accuracy achieved by the quantum SVM method with the five
different encoding functions.}
\subfigure[Training accuracy]{
\begin{tabular}{ccccc}
       encoding function & Circle & Exp    & Moon   & Xor    \\  \hline
       (\ref{eq:ef1})    & 1.00   & 0.91   & 0.85 & 1.00  \\
       (\ref{eq:ef2})    & 1.00  & 0.93 & 0.96 & 0.97 \\
       (\ref{eq:ef3})    & 1.00 & 0.97  & 0.91 & 0.93 \\
       (\ref{eq:ef4})    & 1.00  & 0.98   & 1.00  & 0.95   \\
       (\ref{eq:ef5})    & 1.00 & 0.94  & 0.98 & 0.93 \\
\end{tabular}
\label{tb:train}
}
\subfigure[Test accuracy]{
\begin{tabular}{ccccc}
       encoding function & Circle & Exp  & Moon & Xor  \\  \hline
       (\ref{eq:ef1})    & 0.97   & 0.83 & 0.85 & 0.99 \\
       (\ref{eq:ef2})    & 0.96   & 0.89 & 0.87 & 0.96 \\
       (\ref{eq:ef3})    & 1.00    & 0.92 & 0.86 & 0.91 \\
       (\ref{eq:ef4})    & 1.00   & 0.88 & 0.92 & 0.89 \\
       (\ref{eq:ef5})    & 1.00   & 0.92 & 0.87 & 0.88 \\
\end{tabular}
\label{tb:test}
}
\end{center}
\end{table}


\subsection{Analysis of the feature map}

\begin{table}
\begin{center}
\caption{The minimum accuracy calculated only with the feature maps and the dataset.}
          \begin{tabular}{ccccc}
             encoding function & Circle & Exp  & Moon & Xor  \\  \hline
             (\ref{eq:ef1})    & 0.99   & 0.77 & 0.83     & 0.99 \\
             (\ref{eq:ef2})    & 0.99   & 0.76 & 0.80 & 0.91 \\
             (\ref{eq:ef3})    & 0.99   & 0.86   & 0.89 & 0.85 \\
             (\ref{eq:ef4})    & 0.99   & 0.88    & 0.89     & 0.84 \\
             (\ref{eq:ef5})    & 0.99   & 0.81 & 0.85 & 0.78 \\
          \end{tabular}
\label{tb:MinimumAccuracy}
\end{center}
\end{table}

\begin{figure}[tb]
\begin{center}
\includegraphics[width=1\linewidth]{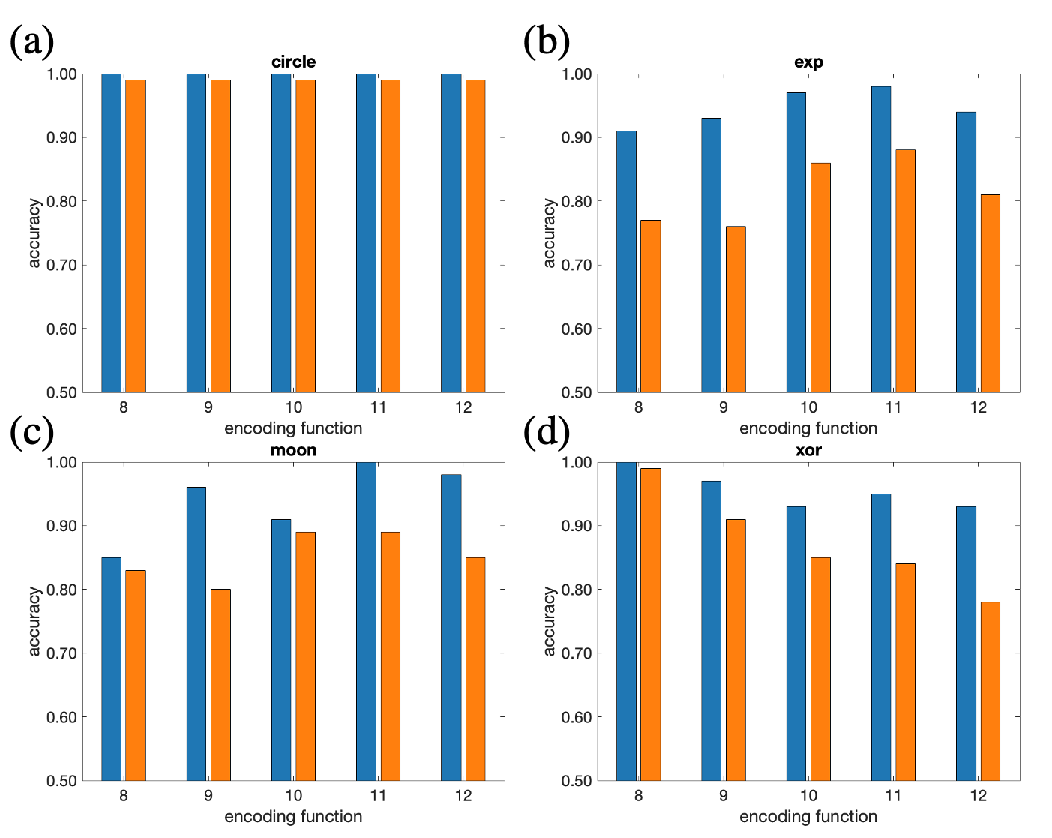}
\caption{Comparison between the exact training accuracy in Table~\ref{tb:train} (blue)
and the minimum accuracy in Table~\ref{tb:MinimumAccuracy} (orange).}
\label{fig:compareTables}
\end{center}
\end{figure}

Here we examine, for the classification problem described above, how the feature map
$\bm{a}(\bm{x})$ would benefit us to figure out the distribution of dataset in the feature
space and whether the minimum accuracy would actually predict a suitable encoding
function.

\begin{figure}[tbp]
\begin{center}
\includegraphics[width=0.8\linewidth]{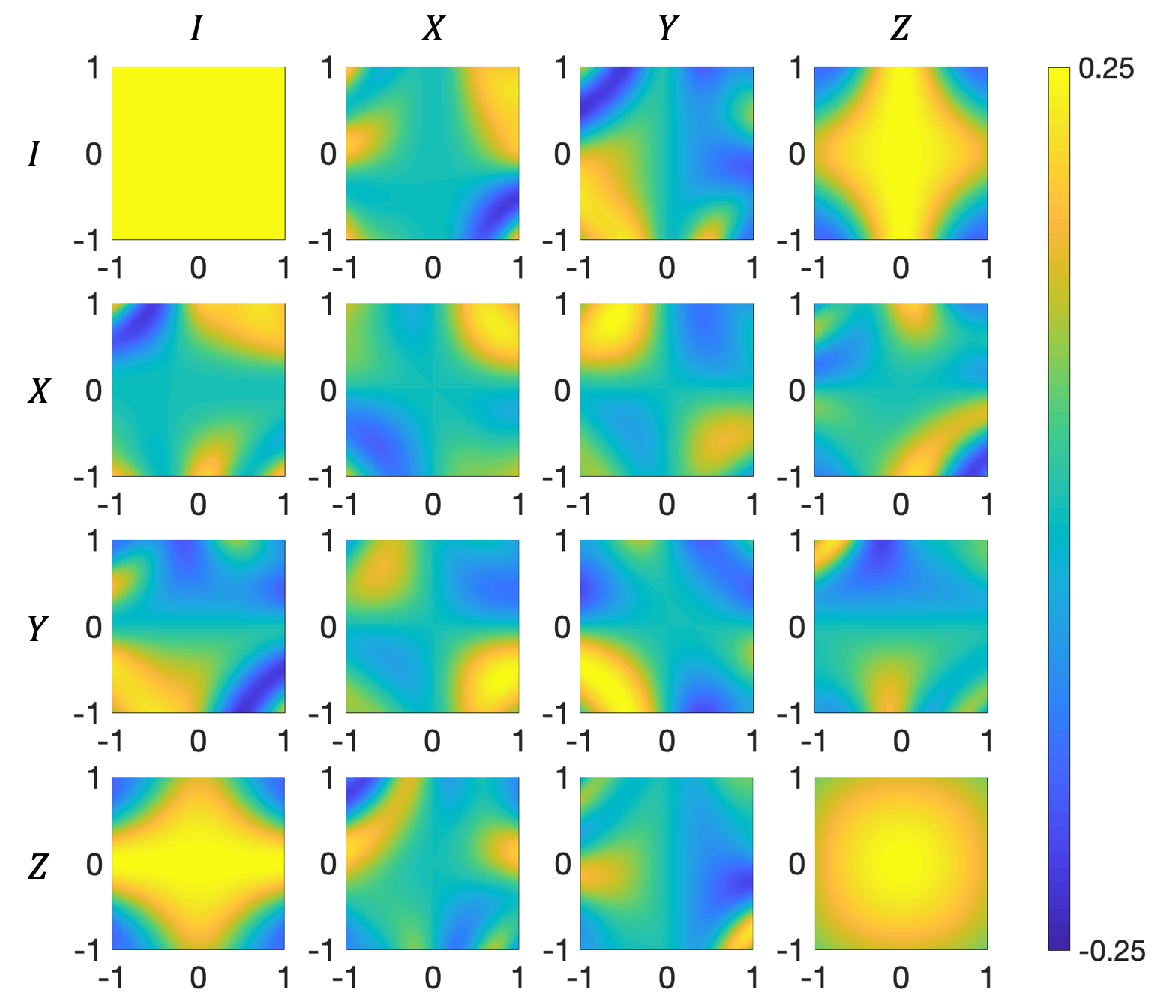}
\caption{Color map of all the 2-dimensional spaces $a_i(\bm{x})$ with the encoding
function \eqref{eq:ef1}.}
\label{fig:ef1}
\end{center}
\end{figure}

\begin{figure}[tbp]
\begin{center}
\includegraphics[width=0.8\linewidth]{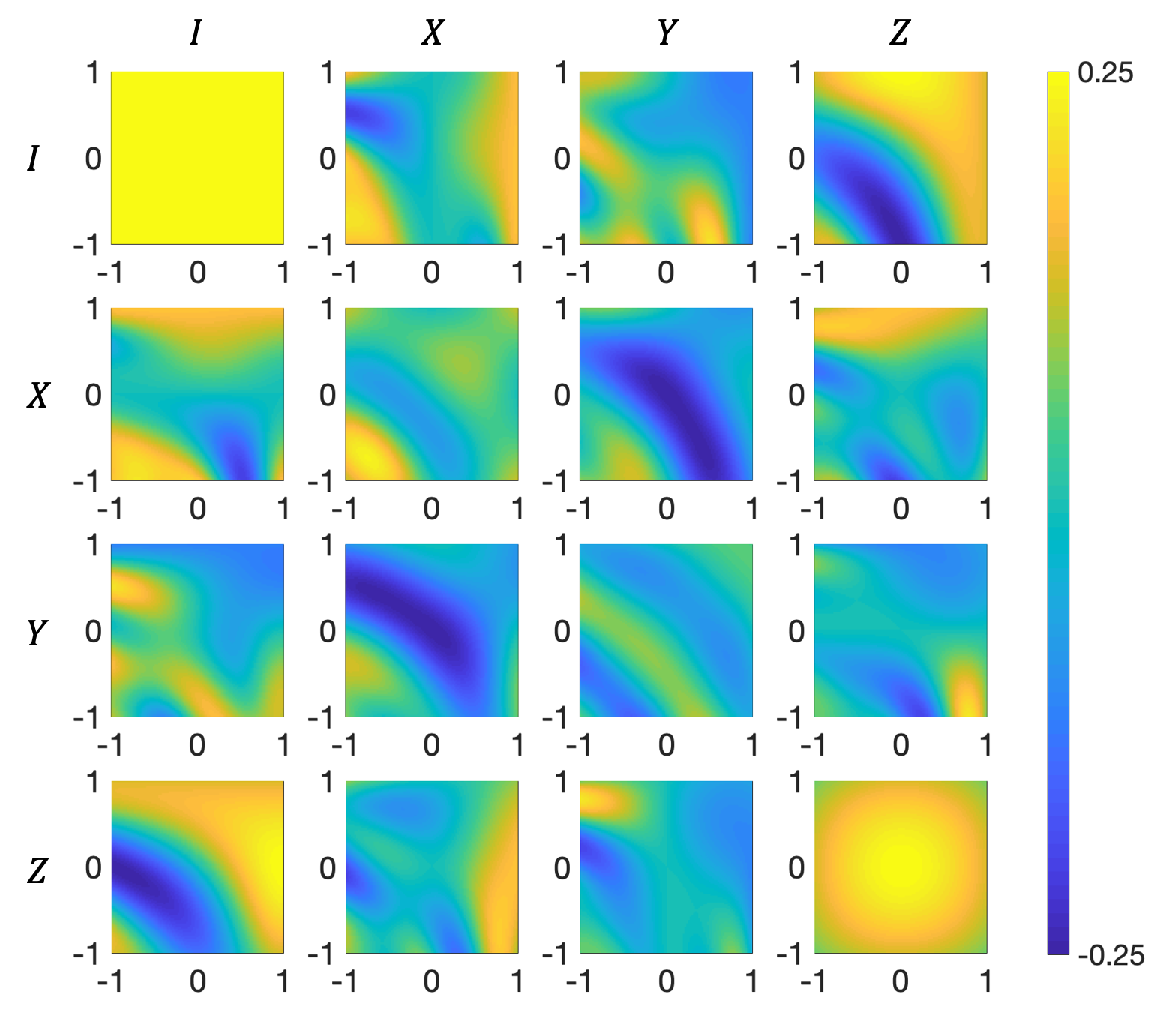}
\caption{Color map of all the 2-dimensional spaces $a_i(\bm{x})$ with the encoding
function \eqref{eq:ef2}.}
\label{fig:ef2}
\end{center}
\end{figure}

\begin{figure}[tbp]
\begin{center}
\includegraphics[width=0.8\linewidth]{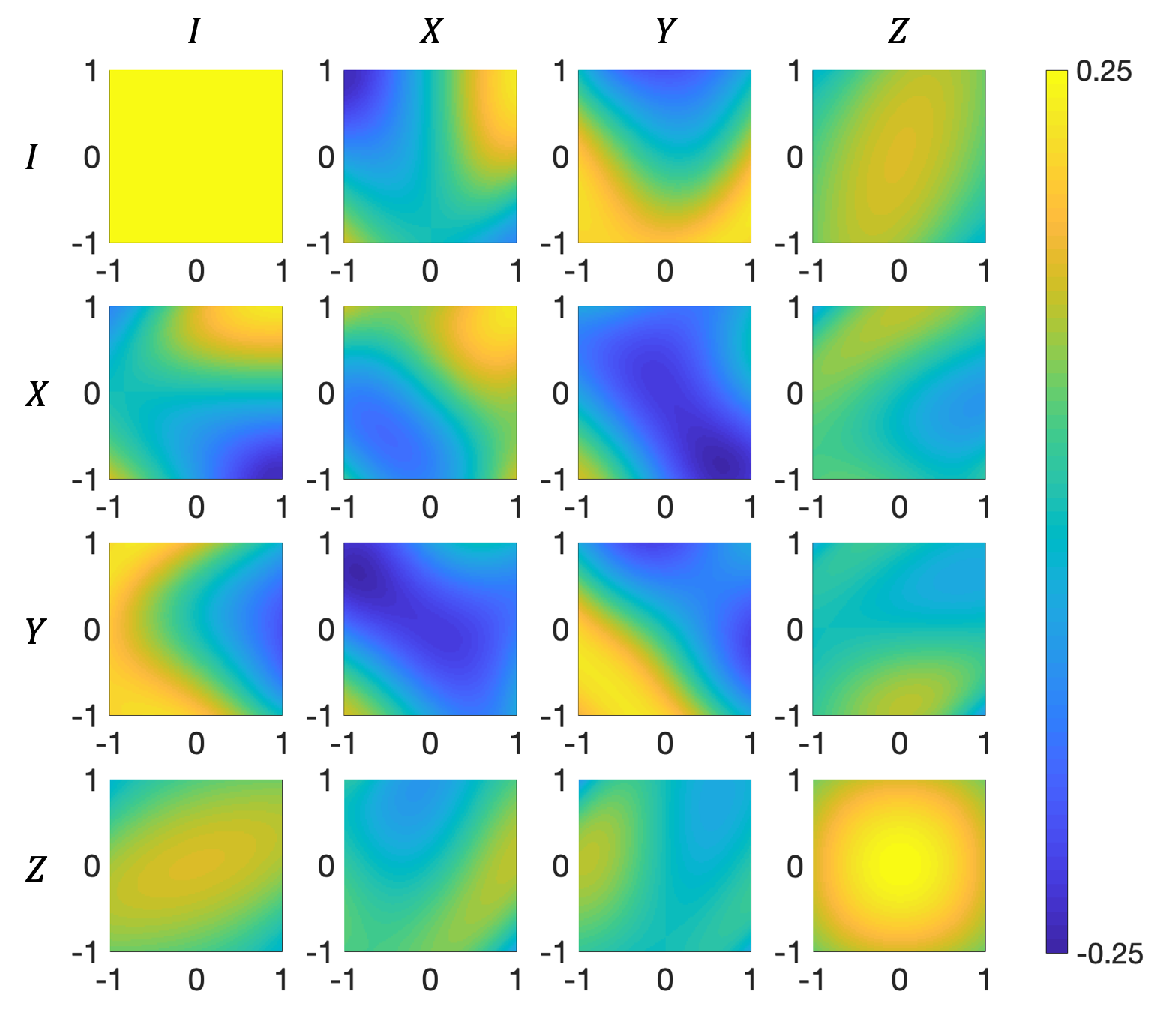}
\caption{Color map of all the 2-dimensional spaces $a_i(\bm{x})$ with the encoding
function \eqref{eq:ef3}.}
\label{fig:ef3}
\end{center}
\end{figure}

\begin{figure}[tbp]
\begin{center}
\includegraphics[width=0.8\linewidth]{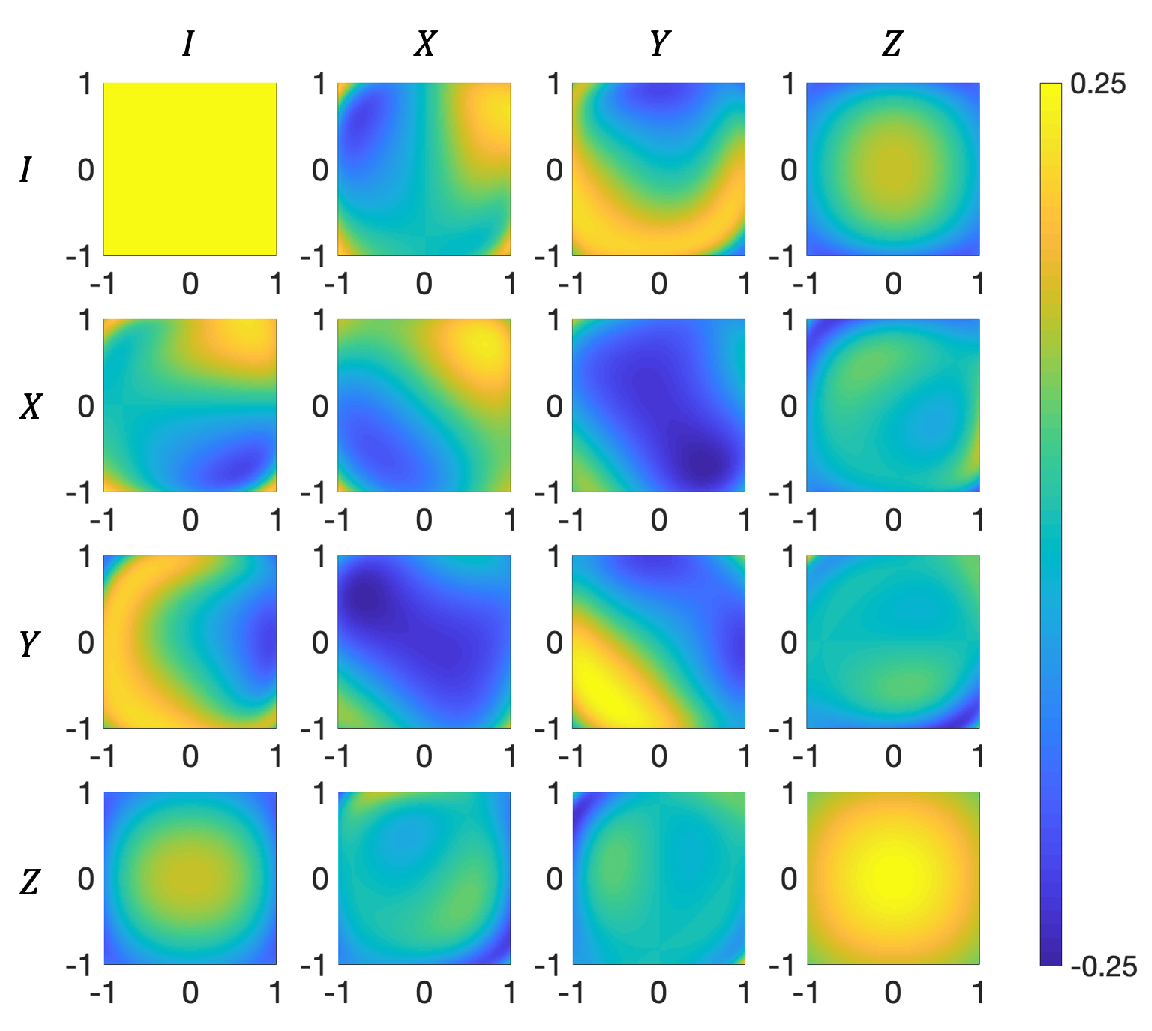}
\caption{Color map of all the 2-dimensional spaces $a_i(\bm{x})$ with the encoding
function \eqref{eq:ef4}.}
\label{fig:ef4}
\end{center}
\end{figure}

\begin{figure}[tbp]
\begin{center}
\includegraphics[width=0.8\linewidth]{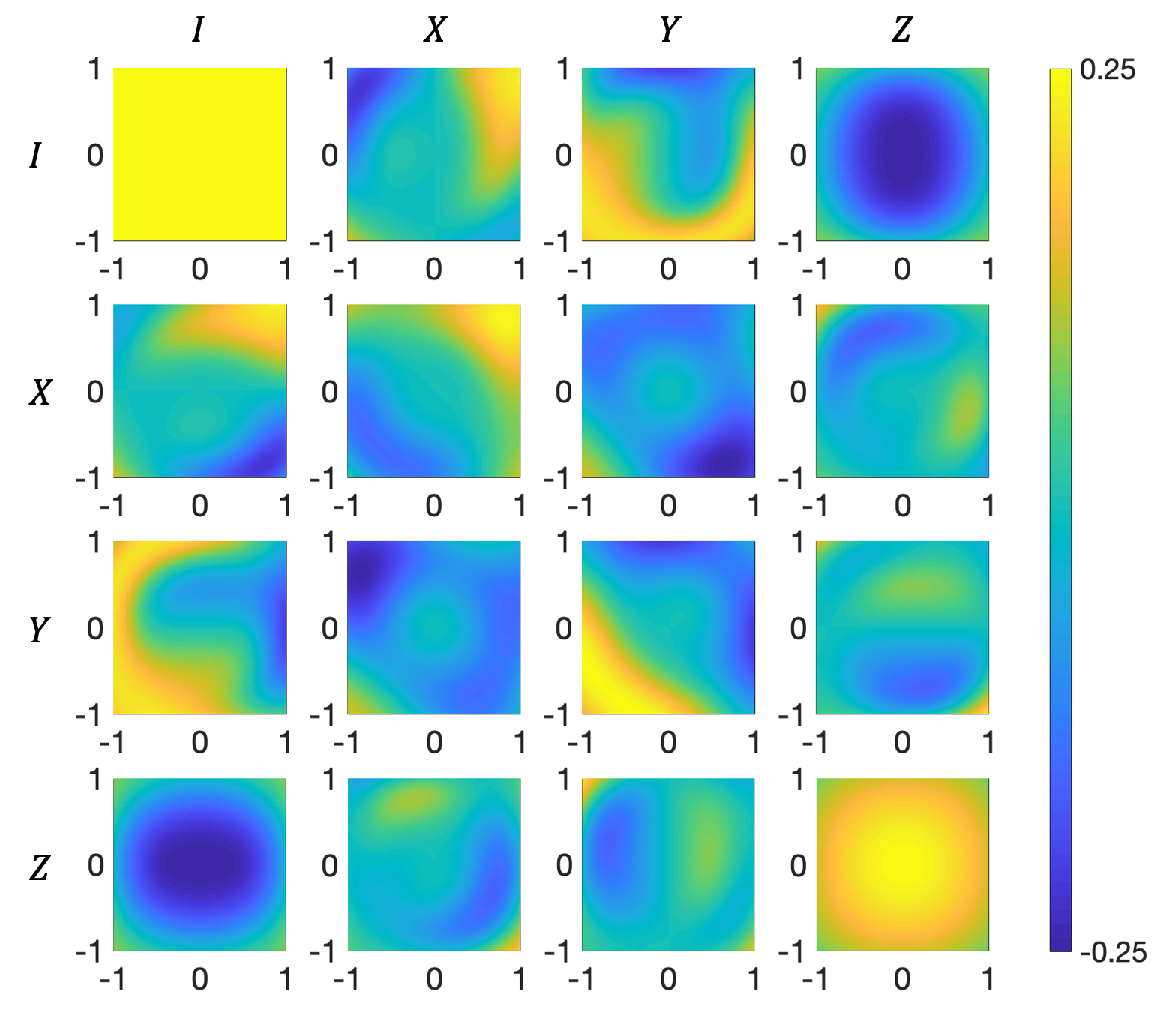}
\caption{Color map of all the 2-dimensional spaces $a_i(\bm{x})$ with the encoding
function \eqref{eq:ef5}.}
\label{fig:ef5}
\end{center}
\end{figure}

First, Figs.~\ref{fig:ef1}-\ref{fig:ef5} show the color map of $a_i(\bm{x})$ listed in
Table~\ref{tb:pdc}, as a function of $\bm{x}\in\mathbb{R}^2$ with $i=II, XI, YI, \ldots, ZZ$
for the encoding functions \eqref{eq:ef1}-\eqref{eq:ef5}, respectively.
Note that each $a_i(\bm{x})$ is not determined from a specific input dataset.
Nevertheless, very interestingly, some of those 2-dimensional spaces intrinsically possess
the shape of distribution of the coming input dataset, which will thus affect on the
resulting classification accuracy.
For example, in all cases of Figs.~\ref{fig:ef1}-\ref{fig:ef5} the ZZ element $a_{ZZ}(\bm{x})$
has a circle shape, meaning that Circle dataset can be classified only by $a_{ZZ}(\bm{x})$;
this observation is consistent with the fact that Circle dataset can be indeed classified with
high training/test accuracies as shown in Tables \ref{tb:train} and \ref{tb:test}.
Similar results can also be clearly found in Fig.~\ref{fig:ef1} (the case of encoding function
\eqref{eq:ef1}) and in Fig.~\ref{fig:ef4} (the case of \eqref{eq:ef4});
the shape of $a_{YX}(\bm{x})$ in Fig.~\ref{fig:ef1} has a similar distribution to Xor dataset,
and actually \eqref{eq:ef1} achieves the best training accuracy 1.00 for Xor dataset;
the shape of $a_{YI}(\bm{x})$ in Fig.~\ref{fig:ef4} has a similar distribution to Exp dataset,
and actually \eqref{eq:ef4} achieves the high training accuracy 0.98 for Exp dataset.

Next, Table~\ref{tb:MinimumAccuracy} gives the minimum accuracy for the encoding
functions \eqref{eq:ef1}-\eqref{eq:ef5}, which are calculated according to the procedure
given in Section~\ref{sec:min accuracy}.
Recall that the minimum accuracy gives a lower bound of the exact training accuracy
achieved by any optimized classifier, or in other words, it guarantees a worst-case
accuracy.
Hence, the minimum accuracy may be used as a guide to determine the feature map;
that is, the encoding function with the largest minimum accuracy is recommended.
Then, for Moon dataset, Table~\ref{tb:MinimumAccuracy} suggests the encoding functions
\eqref{eq:ef3} or \eqref{eq:ef4};
similarly, for Exp and Xor datasets, \eqref{eq:ef4} and \eqref{eq:ef1} are recommended,
respectively.
Also, for the case of Circle dataset, any encoding function can be used.

Now let us compare the minimum accuracy with the exact training accuracies given in
Table~\ref{tb:train}, to see if the above suggestions are consistent to the actual classification
performance achieved by the quantum SVM.
Figure~\ref{fig:compareTables} gives the summary, where the minimum and exact accuracies
are indicated with the orange and blue bars, respectively.
Importantly, the encoding function selected according to the aforementioned guide based
on the minimum accuracy produce the best training accuracies;
hence, as expected, the minimum accuracy may be used as a convenient measure for
determining a suitable encoding function and accordingly a good feature map.
Also in many cases we find positive correlation in the minimum accuracy and the exact
accuracy.
In particular we here consider the following simple definition; in each dataset (b), (c), and
(d), a pair of functions are positively correlated if the order of their minimum accuracies is
the same as that of their exact accuracies.
For instance, for all the cases (b), (c), and (d), the function \eqref{eq:ef4} has a higher
minimum accuracy than \eqref{eq:ef5}, and this order holds also for the exact accuracy;
hence they are positively correlated.
In fact, except the pair \eqref{eq:ef3} and \eqref{eq:ef4} in (c), the ratio of positively
correlated functions is $23/29\approx 79\%$.
This fact also supports the validity of the use of minimum accuracy as a reasonable
guide for choosing the encoding function.


\subsection{Synthesis of the feature map via the combined Kernel method:
Toward ensemble learning}

\begin{figure}[tb]
\centering
\includegraphics[width=0.8\linewidth]{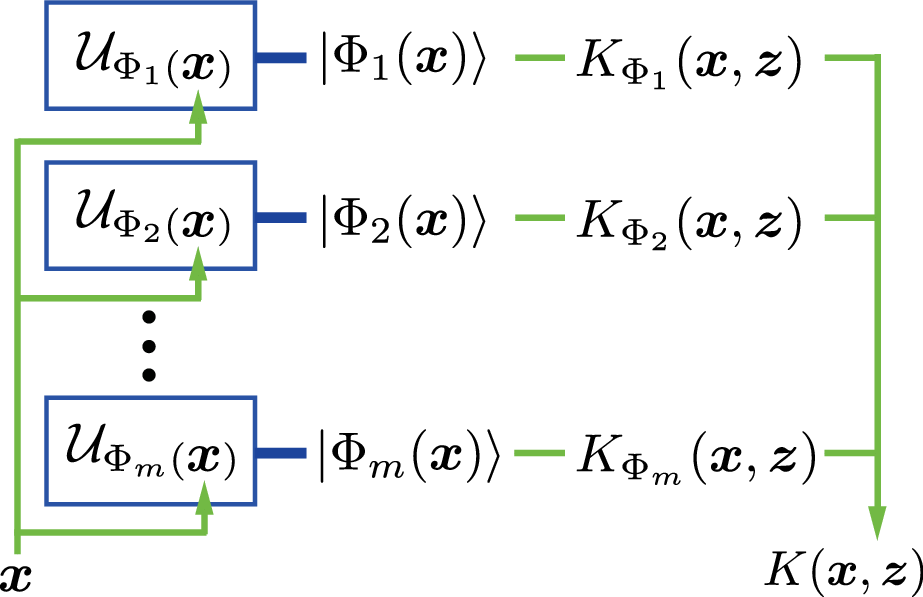}
\caption{
Scheme of the kernel synthesizer composed of different (weak) quantum computers. }
\label{fig:CombKernels}
\end{figure}

\begin{table}
\begin{center}
\caption{Training and Test accuracies via the classifier with combined kernels.}
\subfigure[Classification accuracies for Moon dataset]{
\begin{tabular}{ccccc}
     Encoding function & (\ref{eq:ef1}) + (\ref{eq:ef2}) & (\ref{eq:ef1}) + (\ref{eq:ef3}) & (\ref{eq:ef1}) + (\ref{eq:ef4}) & (\ref{eq:ef1}) + (\ref{eq:ef5})  \\	\hline
     Training    & 1.00   & 0.94  & 1.00   & 1.00 \\
     Test          & 0.95   & 0.90 & 0.98 & 0.96 \\
\end{tabular}
\label{tb:Kernel sum Moon}
}
\subfigure[Classification accuracies for Exp dataset]{
\begin{tabular}{ccccc}
      Encoding function & (\ref{eq:ef1}) + (\ref{eq:ef2}) & (\ref{eq:ef1}) + (\ref{eq:ef3}) & (\ref{eq:ef1}) + (\ref{eq:ef4}) & (\ref{eq:ef1}) + (\ref{eq:ef5})  \\	\hline
      Training    & 0.96   & 0.93  & 0.95   & 0.95 \\
      Test          & 0.92   & 0.90 & 0.88 & 0.92 \\
\end{tabular}
\label{tb:Kernel sum Exp}
}
\end{center}
\end{table}

In the classical regime there have been a number of works on designing efficient kernels;
a simple strategy is to combine some different kernels to construct a single kernel, so that
the constructed one might have a desired characteristic by compensating the weakness
of each kernel \cite{Bishop2006}.
Here we demonstrate that this idea works in quantum regime as well, as actually in the
above sections we have introduced several types of encoding functions which indeed
lead to different kernel functions and accordingly different classification performances.
Note that the idea of combined kernel for quantum classifier was briefly addressed in
\cite{Chatterjee 2017} yet without concrete demonstration.

A typical combining method of kernels is to take a summation of them, as illustrated in
Fig.~\ref{fig:CombKernels}:
\begin{equation*}
     K(\bm{x}, \bm{z}) = \sum_{i = 1}^{m} \lambda_{i} K_{\Phi_i}(\bm{x}, \bm{z}),
\end{equation*}
where $\lambda_i$ are the weighting parameters satisfying $\sum_{i = 1}^{m} \lambda_{i} = m$
with $0\leq \lambda_i \leq m$ (normalization of $\{\lambda_i\}$ does not lead to any essential
difference).
Here, to demonstrate this idea, we consider the combination of two equally-weighted kernels,
i.e., the case of $m=2$ and $\lambda_1=\lambda_2=1$; see \cite{Lanckriet2004,Dios2007}
for the validity of this choice in the classical case.
Even in this simple case a possible advantage may be readily seen;
that is, a sum of kernels theoretically results in a higher dimensional feature space than that
of the original ones as follows:
\begin{align*}
     K_{\Phi_1}(\bm{x}, \bm{z}) +  K_{\Phi_2}(\bm{x}, \bm{z})
      & = |\braket{\Phi_1(\bm{x})|\Phi_1(\bm{z})}|^2 + |\braket{\Phi_2(\bm{x})|\Phi_2(\bm{z})}|^2
\\
      & = |\braket{\Phi_1\oplus\Phi_2(\bm{x})|\Phi_1\oplus\Phi_2(\bm{z})}|^2,
\end{align*}
where $\ket{\Phi_1\oplus\Phi_2(\bm{x})} = \ket{\Phi_1(\bm{x})}\oplus\ket{\Phi_2(\bm{x})}$,
meaning that the classical data $\bm{x}\in\mathbb{R}^{\tilde{n}}$ is encoded into the direct
sum of two Hilbert spaces and hence the dimension of the feature space is doubled.
Here we consider the same benchmark classification problems as above, by applying the
2-qubit classifier with the kernel constructed from the encoding function \eqref{eq:ef1}
and the other four.
In this case $\bm{a}(\bm{x})$ is a 32 dimensional real vector, but with two redundant
elements $a_{II}$ and $a_{ZZ}$.

We first see how much the combined kernel may improve the classification accuracy for
Moon dataset for which the classifier using the single encoding function \eqref{eq:ef1}
showed the worst training accuracy 0.85.
The resultant classification accuracies obtained by applying the combined kernels are shown
in Table~\ref{tb:Kernel sum Moon}; every kernel results in improving the classification accuracy.
Especially, when combining the weak classifiers with the kernels \eqref{eq:ef1} and
\eqref{eq:ef3}, in which case the training accuracy is 0.85 and 0.91 respectively,
the classifier with this new constructed kernel achieves the accuracy 0.94.
This would make sense, because the feature space visualized by $a_i(\bm{x})$ of the
encoding functions \eqref{eq:ef1} and \eqref{eq:ef3} look very different, indicating that the
advantages of each classifiers might be well synthesized to achieve better classification.

Similarly, we test four combined kernels composed of the encoding function \eqref{eq:ef1}
and the others, to classify Exp dataset for which the single encoding function \eqref{eq:ef1}
led to the worst training accuracy 0.91.
The resultant classification accuracies are shown in Table~\ref{tb:Kernel sum Exp}.
In this situation, however, some of the classification performance were not so improved
compared to the results using the original encoding function.
This issue might be resolved by carefully choosing the weighting parameters $\{\lambda_i\}$
when synthesizing the kernels.

A broad concept behind what we have demonstrated here is the so-called {\it ensemble
learning} \cite{Dietterich2000}, which is a general and effective strategy to combine several
weak classifiers to generate a single stronger classifier.
Actually some quantum extension of this method have been deeply investigated
in \cite{Schuld2018,Ximing2019}.
In our work, each classifier is weak in the sense that their circuit depth and the number
of qubits are severely limited;
also the difference of weak classifiers simply comes from the difference of encoding
functions, and the single stronger classifier is constructed merely by taking the summation
of the corresponding kernels.
Systematic strategy for synthesizing weak classifiers for producing a single stronger one
is important particularly in the current status where only noisy intermediate-scale quantum
devices are available.


\section{Conclusion}
\label{sec:conclusion}

In this paper, we proposed a method that helps us to analyze and synthesize the feature
map for the 2-qubit kernel-based quantum classifier, based on the real-valued representation
of it;
the minimum accuracy, which serves as a lower bound of the exact accuracy achieved by
any optimized classifier, was introduced as a tool to effectively screen a library of feature
maps suitable for classification;
also the method of combining (weak) feature maps to produce a better-performing map
was demonstrated with some benchmarking classification problems.
It is important to extend the presented method, beyond a demonstration, to a general and
systematic one for constructing a quantum classifier which fully makes use of its intrinsic
power.

We finally remark that, although calculating the minimum accuracy $R = \max_{i} R_i$
$(i=1,\ldots,4^n)$ is intractable when $n\gg 1$, there might be some circumventing
approaches.
For instance, we may take $\underline{R} = \max_{i\in{\cal I}} R_i$ where the elements
of ${\cal I}$ are randomly chosen from all $i\in\{1,\ldots,4^n\}$;
because $\underline{R}$ also serves as a measure to evaluate the worst case accuracy
for any optimized classifier, it is interesting to investigate how to construct ${\cal I}$
to have a good measure while keeping the size of ${\cal I}$ tractable.
Another interesting direction is to study the connection to the quantum random access
coding \cite{Ambainis,Iwama}, which discusses the method encoding large-size classical
bits into small-size quantum bits;
hence it is expected that even a relatively small-size quantum classifier such that $R$
can be calculated in a reasonable time might have some quantum advantages.
We will work out these problems in the future.

\begin{acknowledgements}
This work was supported by MEXT Quantum Leap Flagship Program
Grant Number JPMXS0118067285 and Cabinet Office PRISM.
\end{acknowledgements}

%
%



\end{document}